# Active beam steering and afocal zooming by nematic liquid crystal infiltrated graded index photonic structures

Ceren Babayigit[1,*], Hamza Kurt[1] and Mirbek Turduev[2]

[1]Department of Electrical and Electronics Engineering, TOBB University of Economics and Technology, Ankara 06560, Turkey.
[2]Department of Electrical and Electronics Engineering, TED University, Ankara 06420, Turkey.
*e-mail: cbabayigit@etu.edu.tr

*Abstract* — This study presents active beam steering and afocal zooming of light by incorporating liquid crystals (LCs) with graded index photonic crystal (GRIN PC). The GRIN PC structures are composed of low refractive index polymer annular rods having gradually varied radii of holes. To actively manipulate incident light, the annular rods are infiltrated with nematic LCs. By applying an external voltage to the infiltrated LCs, the effective index profile of the low-index GRIN PC structure is modulated without introducing any mechanical movement. The incident beam deflection and corresponding focal distance modulation are tuned only by controlling the applied bias voltage. In the present work, hyperbolic secant refractive index profile is chosen to design GRIN PC structures. To design GRIN PC structure with annular PCs, Maxwell-Garnett effective medium approximation is employed. Moreover, we analytically express the relation between infiltrated LCs and gradient parameter to show the physical background of the tuning ability of the proposed devices. Also, beam steering and afocal zooming devices are analytically investigated via geometrical optics and numerically realized with the help of the finite-difference time-domain method. A beam deflection with an angle change of $\Delta\theta_{out}$= 44° and a light magnification with maximum x2.15 are obtained. LCs are inexpensive materials and work under low voltage/power condition. This feature can be used for designing an electro–optic GRIN PC device that can be useful in various optical applications.

## 1. Introduction

Recent developments in the area of photonic crystals (PCs) have revealed the ability of controlling, manipulating, transmitting and detecting the light by paving the way for the broad range of PC applications, such as waveguides [1], optical switches [2], filters [3], defect-mode micro cavities [4], self-collimation [5] and routing [6] to name a few. On the other hand, although many PC devices have been developed with enhanced and additional optical characteristics, most of them perform only certain functionalities that are predicted by their photonic band gaps or dispersion characteristic within the allowed bands. It means that once they were fabricated, the optical properties of such structures cannot be tuned easily, i.e., these structures can be considered as passive photonic devices. In this regard, not only to enhance the functionality, efficiency, flexibility and compactness but also to reduce the cost of optoelectronic devices; attaining the ability to dynamic control of the optical properties by external parameters becomes crucial. In the last decade, there have been several efforts to achieve optical tunability by using various means such as microfluidics [7], ferroelectric materials [8], graphene [9], dielectric elastomers [10], liquid crystals (LCs) [11], thermal control [12] and mechanical control [13]. Among these techniques, LCs can be considered as a good candidate for optical tunability owing to their low power consumption/dissipation, fabrication cost efficiency and structural compactness [14]. For instance, LCs have been used to tune the band gaps [15]-[18], negative refraction effect [19] in PCs and to design tunable optical waveguides [20, 21]. Moreover, the utilizing of LCs in the designs of diffractive lens, switchable grating, and beam deflector have been proposed in Refs. 22-24.

Light focusing and guiding elements occupy an important role in photonic applications. In order to implement the focusing effect, various notable approaches based on negative refraction [25], Fabry–Perot resonances [26], super-lenses [27], aperiodic metallic waveguide arrays [28], metamaterials [29], photonic nano-jets [30] have been proposed. On the other hand, the light guiding is achieved by using photonic-bandgap systems [31], slot-waveguides [32], plasmonic waveguides [33], coupled resonators [34] and dispersion-based optical routing [35,36] methods. Apart from these techniques, manipulation of light propagation behavior is further enhanced by introducing the concept of gradient-index (GRIN) materials. This powerful tool has been contributed to numerous nano-photonic and optical applications including optical mode coupling [37], focusing [38], light bending [39], wavelength de-multiplexing [40] and mode order converting [41]. Nevertheless, despite the significant progress that has been made in these studies, nearly all of the designed light focusing and light guiding structures operates without tunability characteristic. In other words, the optical properties of these structures are fixed and active control is not possible.

In this study, the computational results are obtained for beam steering and afocal zooming applications. GRIN PC structures are composed by aligned polymer annular rods on air medium and filling them with nematic liquid crystals (LCs). By incorporating LC materials with the GRIN media, the active beam steering and afocal zooming of the incident light are achieved. First, the ray transfer analysis of proposed structures is analytically derived via geometrical optics. Then, the simulations are performed with the 2D finite-difference time-domain (FDTD) method with appropriate boundary conditions. Good agreement between the analytical and numerical methods is observed.

## 2. The problem definition and nematic LCs as a tool for active controlling

Photonic interconnectors are used for efficient light transmission between different optical elements. Here, the main challenging issue is the misalignment between input and output light guiding components. If one considers fiber-optic communication system, misalignment between transmitting and receiving components leads to increase of transmission losses that decrease the maximum data rate relatively. In this case, mechanical alignment of the elements can be quite complicated and inconvenient for durability of the system. For this reason, controlling of the alignment without any displacement of elements by mechanical intervention is crucial. In this case, by using beam steering mechanism this problem can be solved and via this approach, photonic interconnectors can be integrated easily to a system [42]. Optical beam steering is also plays an important role in light detection [43], laser micromachining [44], and microscopically imaging [45-47] applications.

Another active light controlling concept that studied in this work is afocal zooming. The afocal zooming is the important optical effect for microscopy and imaging systems [48]. Optical lens system can focus, diverge and collimate the incoming light. Here, for focusing and diverging behaviour of light, imaging properties of the lens system can be modelled via the focal length characteristic. On the other hand, if the focal length of collimated light is infinite, then paraxial characteristic of the lens system cannot be modelled. Such systems are referred as afocal lens systems [49]. Afocal lens systems are mostly used in optical zooming applications. Optical zooming can be obtained by using mechanically compensated or optically compensated zoom lenses that are used for beam expansion and compression purposes [50]. Here, optically compensated lens systems depend on the single motion of a lens or lens system while mechanical compensated zoom lenses operate with multiple motions in multiple directions. Both zooming approaches require complicated and simultaneous adjustment of multiple lens pairs. For this reason, employing these systems in micro/nano photonic applications can be considered as a challenging task. Nevertheless, in order to solve this issue, the monolithic system with a cascaded construction is proposed [51]. However, for different magnification cases, distinct monolithic systems are required, which makes this approach unfeasible.

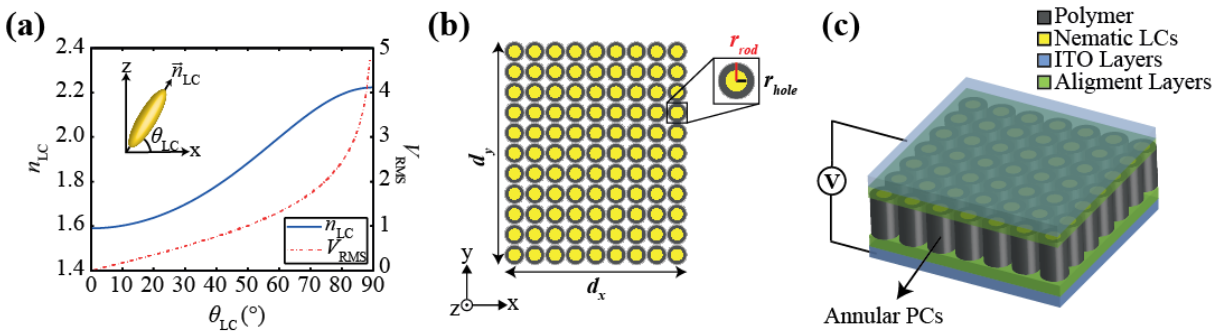

Fig. 1. (a) Corresponding refractive index alteration with respect to $\theta_{LC}$ and the relationship between the rotation angle and applied voltage $V_{RMS}$. (b) Schematic views of the 2D LC-infiltrated annular PC structure and (c) its envisioned voltage excitation mechanism (3D cross-sectional side view).

In this study, we propose beam steering and afocal zooming concepts using nematic LCs infiltrated PC structures. Here, active controlling of the beam deflection and spot size tuning is achieved by applying external voltages. Nematic LCs are birefringent materials which provides different refractive indices along different optical axes [52, 53]. When the ray of light is incident upon a birefringent medium, it splits into two mutually perpendicular components which are called the ordinary rays (o-rays) and extraordinary rays (e-rays). The o-ray always lies perpendicular to the optical axis while e-ray lies in a plane of the optical axis [53]. This phenomenon brings out ordinary ($n_o$) and extraordinary ($n_e$) refractive indices for nematic LCs.

Hence, orientation of the nematic LCs can be tuned by applying external voltage and consequently the effective refractive index of the nematic LCs can be altered [54]. For the transverse magnetic (TM) polarized light where the concerned non-zero electric and magnetic field components are $E_z$, $H_x$, and $H_y$, respectively, the effective permittivity of nematic LCs can be expressed as [55]:

$$\varepsilon_{LC} = n_{LC}^2 = \frac{n_e^2 n_o^2}{n_e^2 \cos(\theta_{LC})^2 + n_o^2 \sin(\theta_{LC})^2} \qquad (1)$$

where $\theta_{LC}$ is the rotation angle of nematic LCs (the nematic LC molecules' orientation along $xz$-plane is given as an inset in Fig. 1(a)). Since the rotation angle $\theta_{LC}$ defines the variation of effective refractive index of nematic LCs, it can be expressed as a function of applied external voltage as follows:

$$\theta_{LC} = \begin{cases} 0 & V \leq V_C \\ \frac{\pi}{2} - 2\tan^{-1}\exp\left(-\frac{V-V_C}{V_0}\right), & V > V_C \end{cases} \qquad (2)$$

Here, the nematic LC molecules can be oriented by an electric field (or a magnetic field) when the field strength exceeds the Freedericksz transition threshold [56]. In Eq. 2, $V_c$ is the critical voltage at which the tilting angle starts to change and $V_0$ is a constant value. Also, the term ($V-V_c/V_0$) can be defined as root-mean-square (RMS) voltage, $V_{RMS}$. As a result, corresponding relationships of nematic LCs' the rotation angle with respect to refractive index alteration and $V_{RMS}$ are represented in Fig. 1(a). In Fig. 1(b) the schematic representation of square lattice PC structure infiltrated by nematic LCs is presented. In order to excite nematic LCs by an external voltage a pair of indium tin oxide (ITO) conducting glass substrates are generally used in the way that given in Fig. 1(c). As it is seen in Fig. 1(c), in addition to ITO layers, the "alignment layers" are also used to provide the specific initial angle of nematic LCs' orientation. This initial angle is crucial to obtain a defect-free alignment and to improve the performance of nematic LCs (response time and viewing angle). Another important parameter that influences the performance of the nematic LCs is the temperature. According to Clausius-Mossotti and Lorentz-Lorenz equations the temperature dependency of the nematic LCs' birefringence is expressed as follows [57]:

$$\Delta n = \Delta n_0 \left(1 - \frac{T}{T_C}\right)^\beta, \qquad (3)$$

where, T is the operating temperature, $T_c$ is the clearing temperature of the nematic LC material, β is an exponent and $\Delta n_0$ stands for the nematic LCs temperature dependent birefringence at T=0 K. Since high birefringence is needed to obtain broad range tunability, throughout this manuscript operating temperature is assumed equal to room temperature. It is important to note that in the case of temperature increase, depending upon the birefringence change, the tuning value will decrease.

## 3. Active beam steering and afocal zooming design approach

As it is well known, the light within the GRIN medium follows curved trajectories because of gradual change in refractive index. Hence, for optical phenomena such as focusing, diverging and collimating the GRIN medium can be used without introducing the curved front and back interfaces for the optical components. In this regard, the main concept of the study lies on merging GRIN optics with nematic LCs to achieve active control of propagation of light. Here, the gradient of the refractive index distribution of GRIN medium can be actively changed to obtain the desired steering and zooming effects.

In order to form GRIN medium, two-dimensional (2D) annular PCs are employed. The substantial design parameters for the annular PCs are the radii of the dielectric rods ($r_{rod}$) and the holes ($r_{hole}$) that drilled into the dielectric rods. By gradually changing air holes' radii one can change dielectric filling ratio of the elementary PC cell, so that GRIN PC medium with desired refractive index profile can be obtained.

In this study, hyperbolic secant (HS) refractive index profile is chosen to design GRIN PC lens. HS refractive index distribution is mathematically expressed as follows:

$$n(y) = n_0 \sec h(\alpha y), \qquad (4)$$

where $n_0$ is the refractive index value at the optical axis and $\alpha$ is the gradient parameter that represents the depth of the index distribution. The shape of the HS function closely resembles parabola and by tuning gradient parameter $\alpha$ one can control the sign of the parabola [58]. Here, $\alpha$ parameter can take real and complex values that result in negative (opens downward) and positive (opens upward) parabolas, respectively.

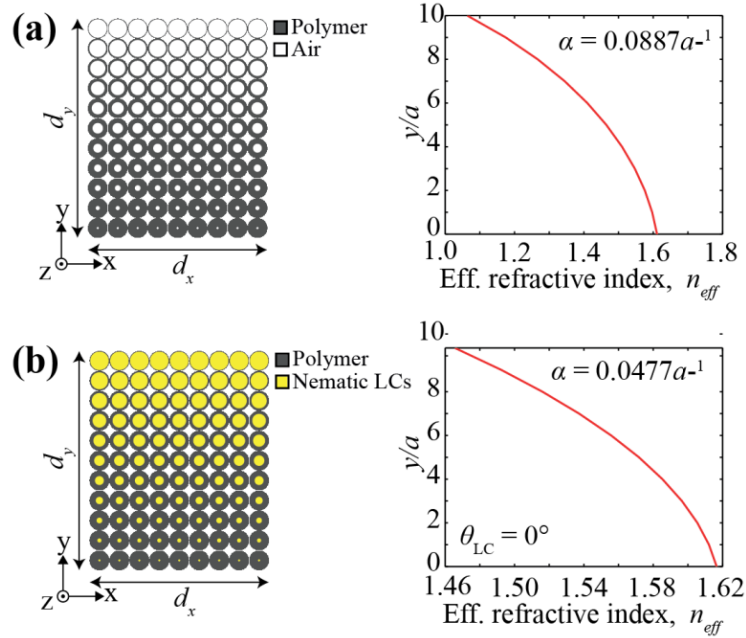

Fig. 2. (a) Schematic of 2D annular GRIN PC structure with varying gradually changing air holes' radii and its corresponding refractive index profile. (b) Schematic of 2D GRIN LC-infiltrated annular PC structures and its corresponding refractive index profile.

In order to design GRIN PC structure with annular PCs, Maxwell-Garnett EM approximation is employed. Here, EM approximation determines the radii of the air holes according to the defined HS refractive index profile. The formulation of the effective permittivity for the transverse magnetic (TM) polarized wave is expressed as follows:

$$\varepsilon_{eff} = \varepsilon_{rod}(f) + \varepsilon_{air}(1-f), \quad (5)$$

where $f = \pi(r_{rod}^2 - r_{hole}^2)/a^2$ is the filling dielectric filling ratio of annular PC rods. Also, $\varepsilon_{rod}$ and $\varepsilon_{air}$ are correspond to the permittivity of annular PC rod and air medium. Then, after some algebra, the variation formula of air holes' radii $r_{hole}$ for TM polarization can be expressed as follows:

$$r_{hole}(y) = \sqrt{r_{rod}^2 + \frac{a^2}{\pi}\frac{(\varepsilon_{air} - n_{eff}^2(y))}{(\varepsilon_{rod} - \varepsilon_{air})}}, \quad (6)$$

where "$a$" is the lattice constant and $n_{eff}(y)$ stands for HS refractive index function. It should be noted that the gradual variation of the air holes is achieved only along the transverse $y$-direction and the positions of the annular PCs cell are kept constant as unit distance $a$ along the y- and x-directions. Herewith, by using Eqs. 5 and 6, one can design the square lattice GRIN PC structure for the desired refractive index distribution. For the design of GRIN PC structure, annular rods are assumed to be made of polymer with refractive index of $n_p$=1.80 [59] and the outer radii are fixed to $r_{rod}$=0.48$a$. The detailed schematic of the designed annular GRIN PC and its effective index profile are depicted in Fig. 2(a). Here, the gradient parameter is calculated as $\alpha$=0.0887$a^{-1}$.

In order to design the tunable GRIN PC structure, the air holes of annular PCs are assumed to be infiltrated by nematic LC molecules with $n_o$=1.59 and $n_e$=2.22. Here, refractive indices correspond to phenylacetylene type LC material [60]. Infiltration by nematic LC leads to change in effective refractive index distribution of the annular GRIN PC structure according to the following formula:

$$n_{eff}^2 = \varepsilon_{eff} = \varepsilon_{rod}(ff_2 - ff_1) + \varepsilon_{LC}(ff_1) + \varepsilon_{air}(1 - ff_2), \quad (7)$$

where $\varepsilon_{LC}$ is the permittivity of the nematic LCs, $ff_1 = \pi r_{hole}^2/a^2$ and $ff_2 = \pi r_{rod}^2/a^2$ are the filling factors of the holes area that drilled into the dielectric rods and annular PC rods, respectively. The schematic representation of the LC infiltrated structure with its corresponding refractive index profile is given in Fig. 2(b). In this case, the gradient parameter $\alpha$ is recalculated as 0.0477$a^{-1}$. As can be seen from the index profiles in Figs. 2(a) and 2(b), one can observe that infiltration of LCs to the annular GRIN PC structure affects gradient parameter. In order to analytically express the relation between infiltrated LC and $\alpha$ parameter, the following formula can be obtained by incorporating Eqs. 4 and 7:

$$\alpha = \frac{1}{y}\sec h^{-1}\left(\frac{1}{n_0}\sqrt{\frac{\pi r_{hole}^2(y)}{a^2}(\varepsilon_{LC}-\varepsilon_{rod})+\frac{\pi r_{rod}^2}{a^2}(\varepsilon_{rod}-\varepsilon_{air})+\varepsilon_{air}}\right). \qquad (8)$$

In Eq. 8 all parameters except $\varepsilon_{LC}$ are defined by constant values so that $\alpha$ depends on only $\varepsilon_{LC}$ which varies with the rotation angle of nematic LCs, $\theta_{LC}$. The refractive index profile variation of annular GRIN PC with respect to the rotation angle $\theta_{LC}$ of the infiltrated nematic LCs is presented in the Fig. 3(a). Here, one can see the transition between positive and negative paraboloid index profiles. This transition is related to the values of $\alpha$ parameter, which takes real and imaginary values (this can be proved also by the Eq. 8). As can be seen in Fig. 3(b), for the variation in $\theta_{LC}$ from 0° ($\varepsilon_{LC}$=1.59) to 43° ($\varepsilon_{LC}$=1.80), the gradient parameter gets real values and it decays to the zero with the increase of the rotation angle. On the other hand, $\alpha$ takes imaginary values for the rotation angles $\theta_{LC}$ greater than 43° as shown in Fig. 3(c). Here, the important point is that for $\theta_{LC} = 43°$ the refractive index profile close to the straight line (constant along y-axis) and the $\alpha$ parameter approximately equals to zero.

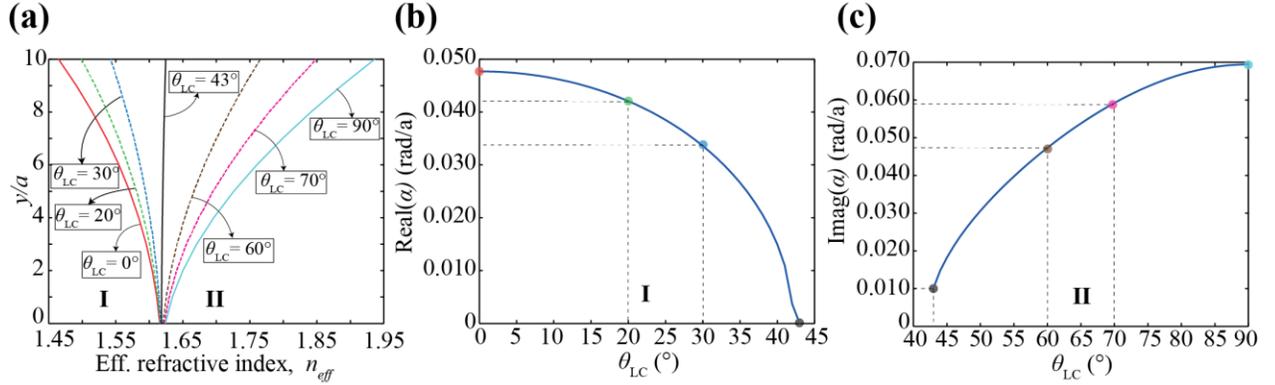

Fig. 3. (a) The effective refractive index change through the y-axis with respect to the LCs' rotation angle. Change of the gradient parameter α depending on the rotation angle in the range of (b) $\theta_{LC}$=[0°,43°] and (c) $\theta_{LC}$=[43°, 90°].

In light of given gradient parameter tuning concept, we propose beam steering and afocal zooming applications with nematic LCs infiltrated annular GRIN PC structures. In Fig. 4(a), a perspective view of the proposed beam steering device is presented. As can be seen in Fig. 4(a), we show operating principle of the beam steering device schematically where applied the external voltage tunes the directivity of the incident beam. Indeed, applied voltage changes gradient parameter $\alpha$ by the help of the nematic LCs and consequently paraboloid refractive index profile changes its sign as shown in Fig. 4(b). For the afocal zooming application, the device that designed for beam steering is mirror symmetrized with respect to propagation x-direction. This mirror-symmetry reveals the converging lens characteristics of the design. The proposed afocal-zoom design is represented in Fig. 4(c) where one can see that the device consist of three regions. As can be seen in Fig. 4(c), only the middle section of PC structure is infiltrated with LCs and left as well as right sections are made of non-infiltrated annular GRIN PC structures. Here, the side annular GRIN PCs operate as converging lenses and middle section infiltrated with LCs (shaded in yellow in Fig. 4(c)) serves as the lens with tunable focal point (applied external voltage manipulates focusing characteristic). The principle of beam diameter manipulation can be deduced from variation of effective refractive index profile in Fig. 4(d). Here, when the LCs' orientation is in the range of $0°\leq \theta_{LC}<43°$, the middle section operates as a converging lens. On the other hand, when the LCs' orientation varies as $43°<\theta_{LC}\leq 90°$, the middle section operates as a diverging lens. Consequently, adaptation of the beam diameter to the particular task can be achieved without introducing any mechanical motion of lens such as in Ref. [50].

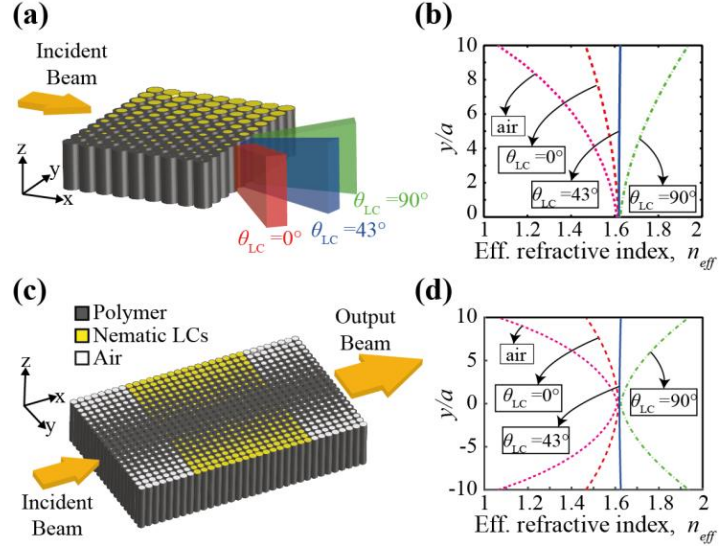

Fig. 4. Perspective view of 2D GRIN LC-infiltrated annular PC structures for (a) beam steering application with the size of $dx=11.96a$ and $dy=9.96a$ and (c) afocal lens structure with the sizes of $dx=32.96a$ and $dy=20.96a$. (b) and (d) show the effective index change through the y-axis with respect to the rotation angle for each configuration.

## 4. Analytical and numerical results

In this section, the operating principle of the proposed beam steering and afocal zooming devices is analytically investigated via ray theory. For this reason, the GRIN media having effective refractive index profiles same as in Figs. 4(c) and 4(d) are examined for beam steering and afocal zooming applications, respectively. After, the time-domain analyses of the proposed structures are provided in detail. Conceptual modelling of light propagation through a non-homogeneous medium can be explained by the ray equation also known as Eikonal equation:

$$\frac{d}{ds}\left[n\frac{dr(s)}{ds}\right]=\nabla n, \qquad (9)$$

where $n$ is the refractive index of medium, $\nabla n$ is the gradient of the $n$, $r(s)=[x(s), y(s)]$ is ray path and $ds=\sqrt{dx^2+dy^2}$ is the arc length along the ray path. By solving the ray equation for given refractive index distribution, ray trajectories can be obtained. In this regard, the ray trajectory function in a continuous HS GRIN medium with refractive index profile given in Eq. 4 is expressed as follows (see Ref. 58 for detailed derivation of the ray trajectory):

$$y(x) = \frac{1}{\alpha}\sinh^{-1}\left(\dot{u}_0 \frac{\sin(\alpha x)}{\alpha} + u_0 \alpha \cos(\alpha x)\right), \tag{10}$$

where $y(x)$ is the ray trajectory function with respect to initial position $u_0$ and incident angle $\dot{u}_0$. Here, for the gradient parameter $\alpha$, the data calculated in Figs. 3(b) and 3(c) are used. It should also be noted that the Eq. 10 is derived in hyperbolic coordinate system by using $u = \sinh(\alpha y)$ transformation. By taking the position derivative of Eq. 10, the slope information of the ray can be expressed as follows:

$$\dot{y}(x) = \frac{-\alpha u_0 \sin(\alpha x) + \dot{u}_0 \cos(\alpha x)}{\alpha \cosh\left(\sinh^{-1}\left[u_0 \alpha \cos(\alpha x) + \dot{u}_0 (\sin(\alpha x)/\alpha)\right]\right)}. \tag{11}$$

By using Eqs. 10 and 11 the ray trajectories through the HS GRIN medium are calculated and the corresponding ray propagations for beam steering application are plotted in Fig. 5. In order to visualize beam steering effect, in Figs. 5(a), 5(b), and 5(c) we depicted three distinct ray deflection cases where gradient parameter equals to $\alpha=0.0477a^{-1}$, $\alpha=i0.0098a^{-1}$, and $\alpha=i0.0691a^{-1}$, respectively. Here, gradient parameters are taken from the data in Figs. 3(b) and 3(c). Here, if $\alpha$ is set to be $0.0477a^{-1} \geq \alpha > 0.036a^{-1}$ (0°≤ $\theta_{LC}$ <42°), the incident beam steers downward (see Fig. 5(a)) and for the interval $0.0144ia^{-1} \leq \alpha \leq 0.0691ia^{-1}$ (44°< $\theta_{LC}$ ≤ 90°) the incident beam steers upwards (see Fig. 5(c)) by the HS GRIN medium. On the other hand, when the gradient parameter $\alpha$ is set to $0.0098ia^{-1}$ ($\theta_{LC}$=43°), the overall medium becomes uniform and thus, incident rays exit the medium without any deflection as shown in Fig. 5(b). The ray deflection angle $\theta_{out}$ variation at the back-surface of the GRIN media with respect to $\alpha$ gradient parameter (LCs rotation angel $\theta_{LC}$) is shown in Fig. 5(d).

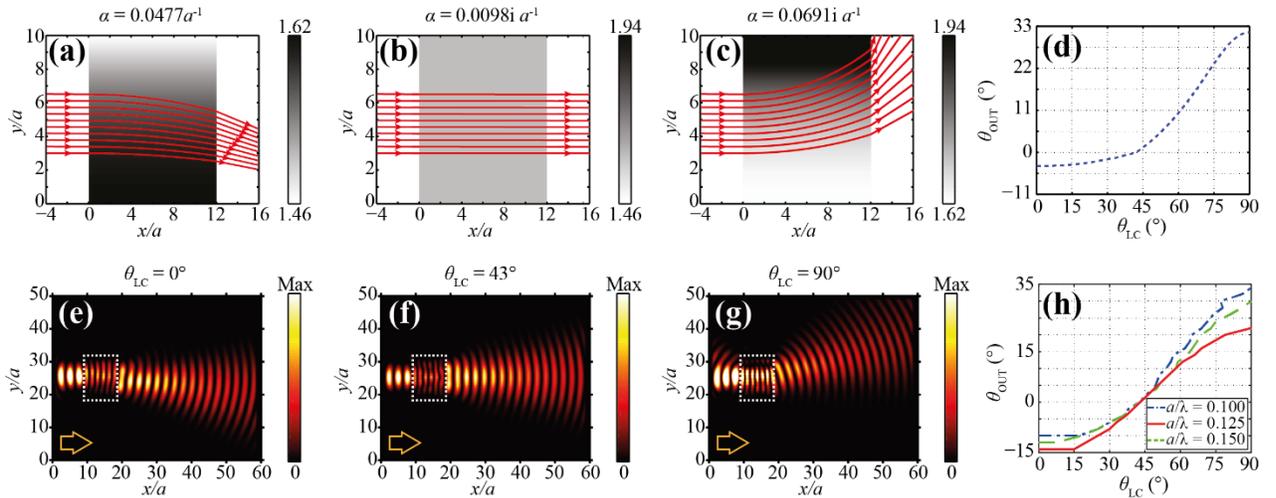

Fig. 5. Expected beam bending based on ray theory in the cases of (a) $\alpha=0.0477a^{-1}$ (b) $\alpha=i0.0098a^{-1}$ and (c) $\alpha=i0.0691a^{-1}$. (d) Beam deflection angle profile in compliance with the rotation angle of the LC molecules. Field intensities for $a/\lambda=0.15$ in the cases of (e) $\theta_{LC} = 0°$, (f) $\theta_{LC} = 43°$ and (g) $\theta_{LC} = 90°$. (h) Analytical beam deflection angle profile with respect to the rotation angle of the LC molecules for $a/\lambda=0.100$, $a/\lambda=0.125$ and $a/\lambda=0.150$.

Table I. Beam steering angle variation with respect to LCs rotation angle $\theta_{LC}$.

| $\theta_{LC}$ | $\theta_{out}$ at $a/\lambda$=0.100 | $\theta_{out}$ at $a/\lambda$=0.125 | $\theta_{out}$ at $a/\lambda$=0.150 |
|---|---|---|---|
| 0° | -10° | -12° | -14° |
| 43° | 0° | 0° | 0° |
| 90° | 34° | 21° | 30° |

After the conceptual explanation of the beam steering application via ray theory, numerical analyses of the proposed LCs infiltrated annular GRIN PC structure are conducted by employing FDTD method. The beam steering structure shown in Fig. 4(a) is exited with TM polarized continuous wave source having a Gaussian profile. Figures 5(e)-5(g) represent the steady state intensity profiles for cases of $\theta_{LC}$ = {0°, 43°, 90°} under incident light wave operating at normalized frequency of $a/\lambda$=0.150. Beam deflection characteristics of the proposed structure can be clearly observed from the figure plots. Also, ray theory observations in Figs. 5(a)-5(c) are verified by the FDTD results (see Figs. 5(e)-5(g)). In order to analyse operating frequency bandwidth of the beam steering PC structure, its steering ability is examined under excitation of light waves with normalized frequencies of $a/\lambda$=0.100, $a/\lambda$=0.125 and $a/\lambda$=0.150 (corresponding to the first TM band). As a result, beam deflection angle variation with respect to LCs rotation angle is presented in Fig. 5(h). The beam deflection results are summarized in Table I where the variation of the steering angle under the effect of the LC orientation angle is presented. As seen from Table I, for the normalized frequencies of $a/\lambda$= {0.100, 0.125, 0.150}, the total deflection angle obtained as $\Delta\theta_{out}$={44°, 43°, 44°} under the rotation angle variation from 0° to 90°.

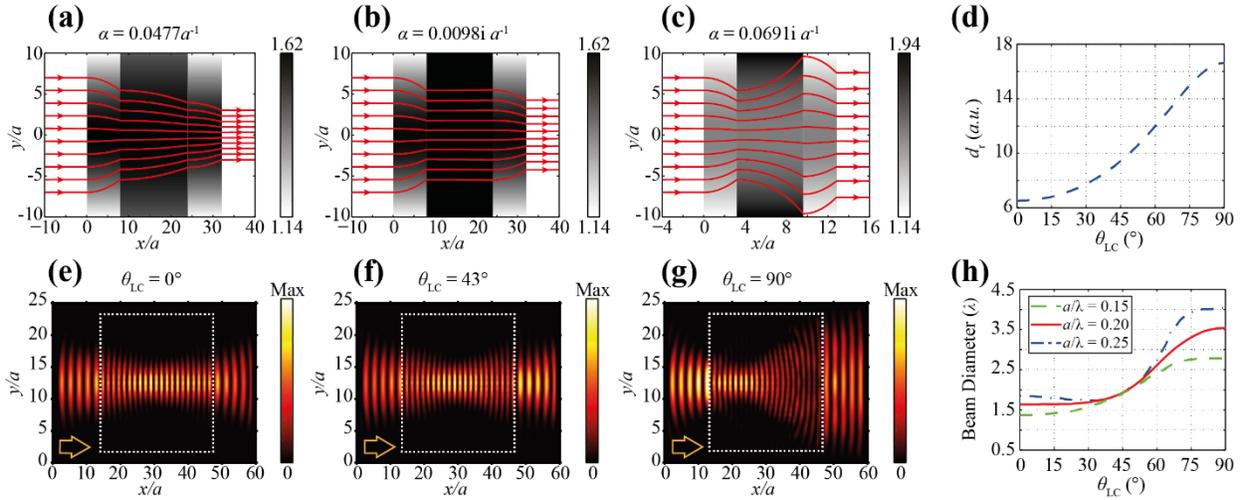

Fig. 6. Expected beam diameter tuning based on ray theory in the cases of (a) $\alpha$=0.0477$a^{-1}$ (b) $\alpha$=i0.0098$a^{-1}$ and (c) $\alpha$=i0.0691$a^{-1}$. (d) Corresponding beam diameter variation of output beams with respect to $\theta_{LC}$. Field intensities for $a/\lambda$=0.20 in the cases of (e) $\theta_{LC}$ = 0°, (f) $\theta_{LC}$ = 43° and (g) $\theta_{LC}$ = 90°. (h) Analytical beam diameter variation of output beams with respect to $\theta_{LC}$ for $a/\lambda$=0.15, $a/\lambda$=0.20 and $a/\lambda$=0.25.

Same procedure is also followed for the proposed afocal zooming application. Similarly, in order to explain the behaviour of light rays within the stacked GRIN media, the ray trajectories are calculated and schematically represented in Figs. 6(a)-6(c). In Fig. 6(a), the exiting positions of propagating rays are accumulated near-by optical axis for the gradient parameter of $\alpha=0.0477a^{-1}$. It means that if one considers a light beam instead of rays, the output beam diameter considered to be squeezed down. On the other hand, if $\alpha$ changes from $\alpha=i0.0098a^{-1}$ (see Fig. 6(b)) to $\alpha=i0.0691a^{-1}$ (see Fig. 6(c)), the locations of exiting rays become distant from each other, i.e., beam diameter expansion characteristics shows up. In order to give an insight to a-focal zooming effect, Fig. 6(d) is prepared. In Fig. 6(d), the $d_r$ distance between upper and lower exiting rays positions with respect to $\alpha$ gradient parameter is represented.

Table II. Beam diemeter variation with respect to LCs rotation angle $\theta_{LC}$.

| $\theta_{LC}$ | $D_{out}$ at $a/\lambda=0.15$ | $D_{out}$ at $a/\lambda=0.20$ | $D_{out}$ at $a/\lambda=0.25$ |
|---|---|---|---|
| 0° | 1.37$\lambda$ | 1.64$\lambda$ | 1.85$\lambda$ |
|  | ($D_{in}$ x 0.73) | ($D_{in}$ x 0.87) | ($D_{in}$ x 0.98) |
| 90° | 2.79$\lambda$ | 3.54$\lambda$ | 4.04$\lambda$ |
|  | ($D_{in}$ x 1.49) | ($D_{in}$ x 1.88) | ($D_{in}$ x 2.15) |

Thereafter, under same simulation conditions given for beam steering device, the numerical analyses of the afocal annular GRIN PC structure is examined via FDTD. Steady state field intensities of the designed zoom lens are determined for $\theta_{LC}= \{0°, 43°, 90°\}$ and the results are shown in Figs. 6(e), 6(f), and 6(g), respectively. As expected, same characteristics which are obtained by the ray theory are corroborated with the numerical results. For the normalized frequencies of $a/\lambda=0.15$, $a/\lambda=0.20$ and $a/\lambda=0.25$, the beam diameter manipulation capacity of the proposed structure is examined in Fig. 6(h). Here, beam diameter calculations are conducted by extracting cross-sectional profiles of the exiting beam at the output of the structure and measuring beam width between the two points where the intensity is $1/e^2= 0.135$ times of the peak value. As a result, beam diameter varies for $a/\lambda=0.15$ from 1.37$\lambda$ to 2.79$\lambda$, for $a/\lambda=0.20$ from 1.64$\lambda$ to 3.54$\lambda$ and for $a/\lambda=0.25$ from 1.85$\lambda$ to 4.04$\lambda$ in the interval of $\theta_{LC}= \{0°, 90°\}$, respectively. The beam diameter tuning results are collected and presented in Table II. Here, the output and input beam diameters are defined as $D_{out}$ and $D_{in}$, respectively.

## 5. Conclusion

In this study, we presented compact structures for tunable beam steering and afocal beam zooming applications by combining LCs and annular PCs. Since there is no moving mechanical part, the proposed structures are more compact and practical when they are compared with their conventional counterparts. The presented calculations show that for different frequency ranges, proposed beam steering and afocal zooming mechanisms supply broad tuning capability. For beam steering application, the deflection angle can be tuned linearly in the range of -12° to 30° by tuning the rotation angle of LCs under the normalized frequency $a/\lambda=0.15$. Similarly, adapting the beam diameter of the output beam from 1.64$\lambda$ to 3.54$\lambda$ is possible with the proposed afocal zooming structure under the normalized frequency $a/\lambda=0.20$. These beam diameters corresponds to x0.87 minimization and x1.88 magnification of the incidence ligth at $a/\lambda=0.20$. As a result, we investigated both analytically and numerically the beam steering and afocal zooming concepts. The designed LCs infiltrated PC structures can be implemented to various optical applications such as near-field scanning, vision correction, object tracking, optical communication, medical diagnostic, x-ray optics, space and atmospheric research, holography and imaging systems.